\documentclass[smallextended]{svjour3}       
\smartqed  
\usepackage{graphicx}
\usepackage{amsmath}
\begin{document}

\title{Implications of a non-zero Poynting flux at infinity sans radiation reaction for a uniformly accelerated charge}
\titlerunning{Implications of Poynting flux for a uniformly accelerated charge}

\author{Ashok K. Singal}
\authorrunning{A. K. Singal} 
\institute{Ashok K. Singal \at
Astronomy and Astrophysics Division, Physical Research Laboratory,
Navrangpura, Ahmedabad - 380 009, India\\
\email{ashokkumar.singal@gmail.com}}
\date{Received: date / Accepted: date}
\maketitle
\begin{abstract}
We investigate in detail the electromagnetic fields of a uniformly accelerated charge, in order to ascertain whether such a charge does `emit' radiation, especially in view of the Poynting flow computed at large distances 
and taken as an evidence of radiation emitted by the charge.
In this context, certain important aspects of the fields need to be taken into account. 
First and foremost is the fact that in the case of a uniformly accelerated charge, one cannot ignore the velocity fields. This then leads to other equally vital points.
The net field energy turns out to be exactly the same as that of a non-accelerated charge having a uniform velocity equal to the instantaneous velocity of the uniformly accelerated charge. 
Further, the Poynting vector, seen with respect to the 'present' location of the uniformly accelerated charge, during the deceleration phase, possesses everywhere a radial component pointing {\em inward} toward the charge, 
becoming nil when the charge becomes momentarily stationary, and during the acceleration phase, 
points away from the charge position. 
Last, but not least, when the leading spherical front of the relativistically beamed Poynting flux, 
advances forward at a large time $t$ to a far-off distance $r=ct$, the charge too 
is not lagging far behind. In fact, these relativistically beamed fields, 
increasingly resemble fields of a charge moving in an inertial frame with a uniform velocity $v_0$, with a convective flow of fields in that frame along with the movement of the charge. 
There is no other Poynting flow in the far-zones that could be termed as {\em radiation emitted} by the charge which, in turn, is fully consistent with the absence of radiation reaction and is also fully conversant with the strong principle of equivalence. 
\end{abstract}
\section{Introduction}
According to Larmor's formula, an accelerated charge emits radiation power proportional to the square of acceleration \cite{1,2,25}, accordingly such radiation should be present in the case of a uniformly accelerated charge too. At the same time, from conservation of energy one expects a proportional radiation reaction on the charge, causing an equivalent loss of power from the  radiating charge. However, it turns out that radiation reaction on a charge is proportional to the rate of change of acceleration \cite{abr05,16,24,3,20,68b}, therefore a uniformly accelerated charge would experience a nil radiation reaction. 
The problem of radiation reaction on the charge due to radiative losses has been extensively discussed in the literature \cite{pa18,51,6,56,57}. On the other hand, if one examines the fields of such a charge in its instantaneous rest frame, there is no magnetic field and thereby no Poynting flux at any distance from the charge, which has given rise to the idea, contrary to the conventional wisdom, that there is no radiation from a uniformly accelerated charge \cite{33}. In fact there is a whole host of arguments in support of the above idea which also  show that neither a freely falling charge in a uniform gravitation field nor a charge supported in such a gravitation field could be emitting electromagnetic radiation \cite{17,18}.

However, if one computes the Poynting flux at a sufficiently large distances from the charge in inertial frames other than the instantaneous rest frame of the charge, then one invariably finds a finite Poynting flux which in literature has been taken as evidence that there {\em is} radiation emitted by a uniformly accelerated charge \cite{5,10,PA02,AL06}.
It has been claimed that the radiation emitted by the uniformly accelerated charge cannot be seen by the co-accelerating observer because the radiation supposedly goes beyond the horizon, into regions of space-time inaccessible to the observer \cite{10,PA02,AL06}. 
The almost paradoxical situation where there is no radiation reaction and hence no energy 
losses by such a charge even though it is supposedly emitting radiation continuously, has sometimes been resolved in the literature by 
proposing the presence of certain Schott-energy term in the near vicinity of the charge that 
presumably helps conserve the overall energy of the charge system and is thought to arise from some sort of acceleration-dependent energy in fields \cite{7,ER002,44,41}. However, of late it has been conclusively demonstrated that the Schott term does not represent an actual energy present in the fields, instead it is merely a difference that shows up in two separate computations of the rate of work done during a real time motion of the charge against the self-force that arises from the time-retarded value of the acceleration of the charge; the two calculations being done with  the velocity and acceleration either both as the retarded time values or both as the real time values \cite{90}. 
In fact, it was explicitly shown that in the case of a uniformly accelerated charge the power said to be radiated, as per Larmor's formula,  which is always calculated in terms of the motion at the retarded-time, is precisely the above difference between the work computations in terms of the present time quantities versus the ones in terms of the retarded time values. 
Further, a critical scrutiny of the electromagnetic fields of a uniformly accelerated charge, where the Schott energy term, due to a relative simplicity of the field expressions should be readily tractable, did not show up anywhere \cite{58b}.

The question that we want to address here is: Does the Poynting flow inferred at infinity \cite{5,10,PA02,AL06} really represents power {\em being emitted} by the uniformly accelerated charge? This is especially relevant in view of an almost universal agreement upon the absence of a radiation reaction on such a charge, in spite of the above Poynting flow.  First we will show that in case of uniform acceleration charge, one cannot ignore the velocity fields. This is because for a constant acceleration $a$, the velocity at the retarded time $t_{\rm r}=-r/c$, comprises a term, $v_{\rm r}\propto at_{\rm r}=-a r/c$, therefore a transverse term of the velocity field, $\propto -v_{\rm r}/r^2= a/cr$, cancels the transverse acceleration field, $\propto -a/cr$, neatly even at large $r$, leaving behind only a transverse field term $\propto v_0/r^2$, where $v_0$ is the instantaneous 'present' velocity of the charge at that instant. The resulting fields are calculated in Sections 2 and 3, which show some quite interesting, but unexpected otherwise, properties leading to a logical resolution of this century-old controversy.

We know that for a charge moving with a uniform velocity, the fields fall with distance as $1/r^2$ and as the charge moves, so do the fields move along with the charge, remaining all around it and following it as if attached to the charge. We call these as self-fields of the charge, and the energy in these fields as the self-field energy. This is in contrast with the free fields, like electromagnetic waves, having independent existence and no longer attached to a charge, even though these might have originated from the charge. Of course there is no radiation from a uniformly moving charge. It is shown in Section 3 that in the case of a uniformly accelerated charge, the net field, including the contributions of velocity fields as well as acceleration fields, falls with distance as $1/r^2$, and behaves quite like the  self-fields of a charge moving with a uniform velocity equal to the instantaneous velocity of the uniformly accelerated charge. 

In Section 4, a computation of the net field energy as well as the net Poynting flux, arising from both acceleration and velocity fields, performed as a function of $r$ with respect to the charge position at the corresponding retarded time, shows that these are exactly the same as found in the self-fields of a charge moving with a uniform velocity equal to the instantaneous velocity $v_0$, of the uniformly accelerated charge. 

Further, in Section 5, it is shown that for a uniformly accelerated charge, moving first in a direction opposite to that of acceleration, during the initial deceleration phase the self-field strength decreases, and the Poynting flow {\em everywhere} is in the {\em inward direction} toward the 'present' charge position, indicating a depletion of energy from the fields. As might be expected, the Poynting flux reduces to a nil value when the charge becomes momentarily stationary with no kinetic energy and a nil transverse field energy; it is only during the acceleration phase, when the charge velocity is increasing, that the Poynting vector points outward from the charge, indicating its increasing self-field strength. That way it is demonstrated how the Poynting flow in question actually goes into building the self-fields of the uniformly accelerated charge with its self-fields growing (reducing) in strength as its velocity increases (decreases) due to its constant acceleration.

Subsequently, in Sections 6 and 7, we also examine in detail the relation between the Poynting flow at a far-off distance $r$ from the charge position at the retarded time $t-r/c$ and the instantaneous charge location at time $t$, when the charge, due to its incessant acceleration, is also moving toward infinity with a velocity approaching the speed of light. In a normal radiation scenario, when the radiation in question is at a large distance ($r\rightarrow \infty$!), the charge responsible is nowhere in the vicinity, assumedly lying not too far from its location at the corresponding retarded time, e.g., in localized charge or current distributions in a radiating antenna. This of course necessarily implies that not only the motion of the charge is bound, with its velocity and acceleration having a sort of oscillatory behaviour, even if not completely a regular harmonic motion. However, for a uniformly accelerated charge, the situation is quite different. At a large $t$, the speed of the uniformly accelerated charge approaches $c$ and it is not far behind the leading spherical front of the Poynting flux advancing with speed $c$ with respect to the time-retarded position of the uniformly accelerated charge, and due to relativistic beaming, lying predominantly in a narrow cone about the direction of motion of the charge. From that 
we shall show that the same relativistically beamed electric field configuration, when considered with respect to the 'present' (instantaneous) position of the now relativistically moving uniformly accelerated charge, closely resembles that of a charge {\em moving uniformly} with a velocity equal to the instantaneous 'present' velocity of the uniformly accelerated charge, 
with no radiation, whatsoever, being `emitted away' by either of them, and naturally, no radiation reaction too in either case. 

\section{Electromagnetic fields of a uniformly accelerated charge in terms of its `real time' motion}
Conventionally, a uniformly accelerated motion is taken to be the one with a constant proper acceleration, where we may assume it to be a one-dimensional motion, since using an appropriate Lorentz transformation, one could reduce the velocity component normal to the acceleration vector to be zero. 

We assume that the charge, coming from $z=\infty$ at time $t=-\infty$, with a velocity approaching $c$ along the $-z$ direction, is under a uniform acceleration, ${\bf a} =\gamma^{3} \dot{\bf v}$,  which is along the $+z$ direction. 
Thus the charge is getting constantly decelerated till at, say, time $t=0$ it comes to rest momentarily  at a location $z=z_0$, for which, without a loss of generality, we can choose the origin of the coordinate system so that $z_0=c^{2}/a$. Afterwards, due to its uniform acceleration, the charge continues to move with an increasing velocity along the $+z$ direction, retracing its earlier path. 

At time $t$, the charge is moving with a velocity $v_0 = c^2t/z_0\gamma_0$, with $\gamma_0 =[{1+(ct/z_0)^2}]^{1/2} $ and occupies a position  $z_{\rm e}=[{z_0^2+c^2t^2}]^{1/2}=z_0\gamma_0$ \cite{88}. Thus during its motion, the charge is at the same location at time $t$ as at $-t$, i.e. $z(t)=z(-t)$, though the velocities at $t$ and $-t$ are equal and opposite, i.e. ${{\bf v}}(t)=-{{\bf v}}(-t)$, with ${{\bf v}}=0$ at $t=0$.  

Due to the cylindrical symmetry of the system, it is convenient to employ  cylindrical coordinates ($\rho,\phi,z$).
Then the electromagnetic field of a uniformly accelerated charge at a point $(\rho,z)$ at time $t$ is given as \cite{5,10,7,88}
\begin{eqnarray}
\label{eq:38ab1}
E_{z}&=&-4ez_0^{2}({z_0^2+c^2t^2}+\rho^{2}-z^{2})/\xi^{3}\nonumber\\
E_{\rho}&=&8ez_0^{2}\rho z/\xi^{3}\nonumber\\
B_{\phi}&=&8ez_0^{2}\rho ct/\xi^{3}\;,
\end{eqnarray}
where $\xi=[(z_0^2+c^2t^2-\rho^{2}-z^{2})^{2}+4z_0^{2}\rho^{2}]^{1/2}$.
The remaining  field components are zero. 

From Eq.~(\ref{eq:38ab1}), we notice that at $t=0$, $B=0$, which means that in the instantaneous rest frame of the charge, the magnetic field is zero everywhere. As first pointed out by Pauli \cite{33} from Born's solutions \cite{32}, this implies that there could be  
no radiation from a uniformly accelerated charge, contradicting Larmor's radiation formula. It has been said that $B=0$ at $t=0$ is unusual for accelerated motion and may be of some interest, but has no bearing on the question of radiation \cite{5,5a}. However, we will show in Section 3 that the magnetic field becomes zero in the instantaneous rest frame of the charge due to a systematic cancellation of the acceleration fields, usually considered to be representing radiation, by the transverse component of the velocity field {\em at all distances} from the charge, thereby implying no radiation.
\begin{figure}[t]
\begin{center}
\includegraphics[width=\columnwidth]{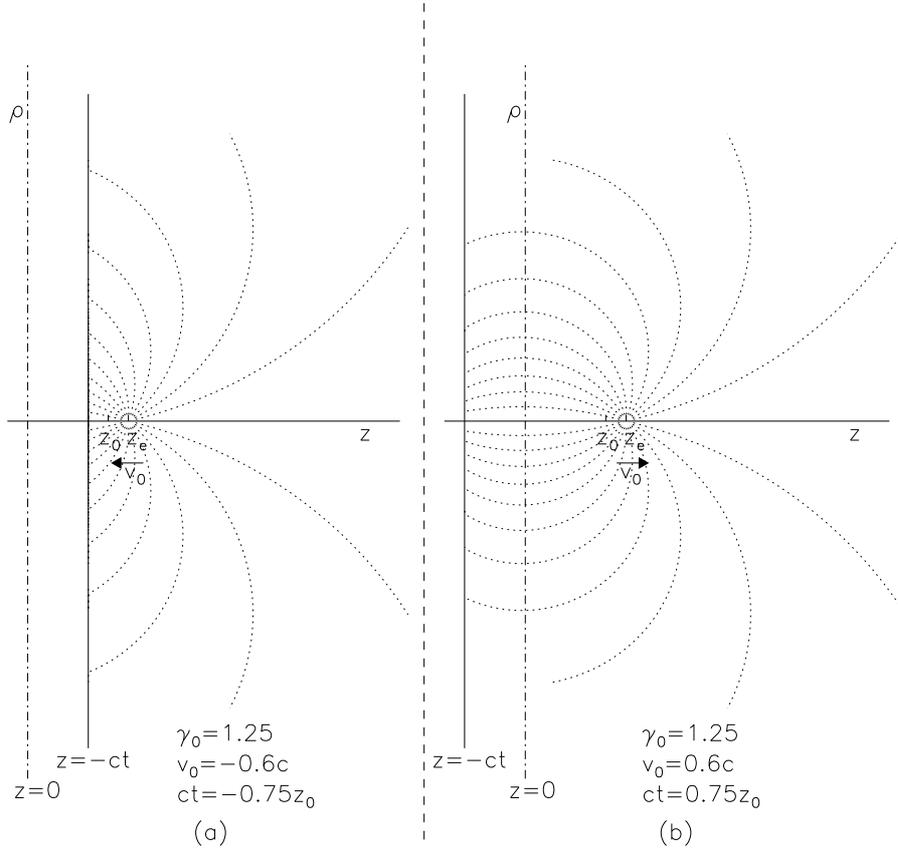}
\caption{Electric field lines of a charge, moving with a uniform proper acceleration along the z-axis. The field lines are drawn for a chosen time (a) $t=-0.75 z_0/c$ when the charge was moving with a velocity $v_0=-0.6c$, (b) at the time $t=0.75 z_0/c$ when the charge was moving with a velocity $v_0=0.6c$. The charge at both events is located at $z_{\rm e}$ and has a corresponding Lorentz factor, $\gamma_0=1.25$. 
The $z=-ct$ plane, in each case, denotes the causality limit of fields.}
\end{center}
\end{figure}
\begin{figure}[t]
\begin{center}
\includegraphics[width=\columnwidth]{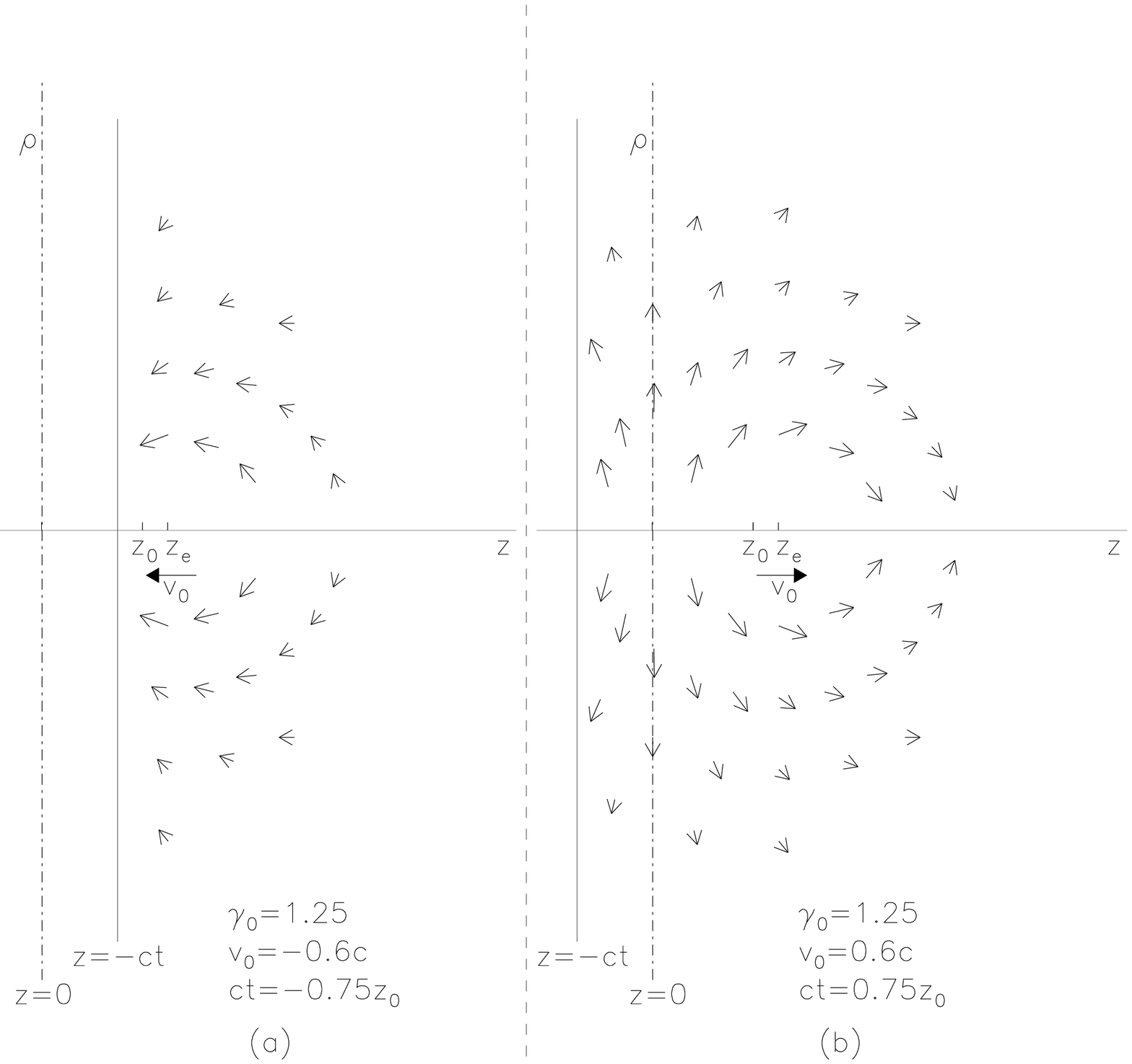}
\caption{Poynting flow around a charge, moving with a uniform proper acceleration along the $z$-axis. Arrows show Poynting vector directions at different distances from the charge, for the two cases shown in Fig.~1. In this schematic diagram, the length of an arrow is not directly proportional to the magnitude of the corresponding Poynting vector, the plot merely indicates the trend qualitatively, with the magnitude of the Poynting vector in successive concentric outer circle falling by a factor of $\sim 10$.}
\end{center}
\end{figure}

Figure~1 shows the electric field lines about the charge location, $z_{\rm e}$, at times $t=\pm t_1$ when the charge was moving with velocity $v_0=\pm ct_1/z_{\rm e}$ for a chosen $ct_1=0.75 z_0$, corresponding to $v_0=\pm 0.6c$ and $\gamma_0=1.25$.  
Figure~2 shows the Poynting flow in the either case.

The $z=-ct$ plane, denoting the  boundary of the causality domain of the field in each case, is also shown in Figs. 1 and 2. At $t<0$, the plane $z=-ct=\gamma_0 z_0 |v_0|/c$, with respect to  the instantaneous position of the charge, lies at a distance $z_{\rm e}-ct$, which for a large $\gamma_0$ is 
\begin{eqnarray}
\label{eq:38ab2}
z_{\rm e}-ct&=&\gamma_0 z_0[1- |v_0|/c]\nonumber\\ 
&\approx& \gamma_0 z_0[1- (1- 1/2\gamma_0^2)]=z_0/2\gamma_0\,, 
\end{eqnarray}
with the plane $z=-ct$ almost passing through the instantaneous charge position. This may of course be understood because for $v_0\approx c$, the ``light-front'' and the charge move almost together. However, due to the deceleration, the charge slows down with time, and at $t=0$, when the charge becomes momentarily stationary at $z_0$, the boundary of the causality domain lies at $z=0$. 
\section{Why is the magnetic field nil in the instantaneous rest-frame?}
Electromagnetic field of a charge $e$, moving with ${\bf v}\parallel \dot{\bf v}$, can be written for any given time $t$ as \cite{1,2,25,28}
\begin{eqnarray}
\label{eq:38aa}
{\bf E}&=&\left[\frac{e({\bf n}-{\bf v}/c)}
{\gamma ^{2}r^{2}(1-{\bf n}\cdot{\bf v}/c)^{3}} +\frac{e\:{\bf n}\times({\bf n}\times
\dot{\bf v})}{rc^2\:(1-{\bf n}\cdot
{\bf v}/c)^{3}}\right]_{t_{\rm r}}\nonumber\\
{\bf B}&=&[{\bf n}] \times {\bf E}
\end{eqnarray}
where quantities within the square brackets are to be evaluated at the corresponding retarded time $t_{\rm r}=t-r/c$. The acceleration field (the second term within the square brackets),  transverse to ${\bf n}$ and falling with distance as $1/r$, is usually assumed to be responsible for radiation from a charge, where the understanding is that the velocity field (the first term within the square brackets) makes only a negligible contribution at large $r$  since it falls with distance as $1/r^2$. 

However, in the case of a uniformly accelerated charge, we will show that a transverse component of velocity fields is not only comparable to but can even cancel the acceleration fields everywhere. In fact, it is this cancellation of the acceleration fields by the velocity  fields that from Eq.~(\ref{eq:38ab1}) had resulted in $B=0$ at $t=0$.  

The total magnetic field, including both the velocity as well as the acceleration field terms from Eq.~(\ref{eq:38aa}), can be written  as 
\begin{eqnarray}
\label{eq:38baa}
\!\!\!\!{\bf B}&=&\left[\frac{-e{\bf n}\times {\bf v}}{\gamma ^{2}r^{2}c(1-{\bf n}\cdot{\bf v}/c)^{3}}+ \frac{-e{\bf n}\times \dot{\bf v}}{r c^{2}(1-{\bf n}\cdot{\bf v}/c)^{3}}\right]_{t_{\rm r}}.
\end{eqnarray}

Now, in Eq.~(\ref{eq:38baa}), $\bf v$ is the velocity at the retarded time $t_{\rm r}=t-r/c$ which, for an assumed one dimensional motion in an inertial frame, with a uniform proper acceleration,  ${\bf a} =\gamma^{3} \dot{\bf v}$, can be written in terms of the instantaneous  velocity ${{\bf v}}_{\rm o}$ of the charge at the `` present'' time $t$ in that frame, as  
\begin{eqnarray}
\label{eq:38c}
[\gamma {\bf v}]_{t_{\rm r}}=\gamma_0 {\bf v}_0- (t-t_{\rm r}){\bf a}=\gamma_0 {\bf v}_0-\left[\frac{r \gamma^{3} \dot{\bf v}}{c}\right]_{t_{\rm r}}\:,
\end{eqnarray}
where $\gamma_0$ is the Lorentz factor at the `` present'' time $t$. 

Substituting Eq.~(\ref{eq:38c}) in Eq.~(\ref{eq:38baa}), we  get the magnetic field of a uniformly accelerated charge as 
\begin{eqnarray}
\nonumber
{\bf B}=\frac{-e\gamma_0{v}_0}{c}\left[\frac{{\bf n}\times \hat{\bf z}}{\gamma ^{3}r^{2}(1-{\bf n}\cdot{\bf v}/c)^{3}} \right]_{t_{\rm r}}\\
\label{eq:38bbb}
=\frac{-e\gamma_0{v}_0}{c}\left[\delta^{3}\frac{{\bf n}\times \hat{\bf z} }{r^{2}} \right]_{t_{\rm r}}\:, 
\end{eqnarray}
where $\hat{\bf z}$ is the unit vector along the direction of motion, say along the $z$-axis and $\delta=1/[\gamma (1-{\bf n}\cdot{\bf v}/c)]$ represents the relativistic beaming, which merely affects the angular distribution of the field around the direction of motion of the charge by $\delta^3$ in that inertial frame.

Thus we find that the magnetic field vector in the case of a uniformly acceleration, at any space-time event, apart from the relativistic beaming factor $\delta^{3}$, is proportional to the value of the instantaneous relativistic velocity ${\gamma_0{v}_0}$ of the charge,  and falls as square of distance ($\propto 1/r^2$), howsoever far it may be from the charge location.
In the instantaneous rest-frame, where, by definition, the present velocity ${{v}_0}=0$, we have a nil magnetic field, ${B}=0$ 
throughout, which means no Poynting flux {\em anywhere}, and consequently no radiation that would be detected by any observer at whatever distance, in the instantaneous rest frame.
This, in turn, is consistent with the expectation that an accelerated charge that is momentarily stationary, has no velocity at that instant and hence no kinetic energy that could be balanced against the radiation losses. 
In fact, as we shall see later, the Poynting flux for a uniformly accelerated charge being zero  in its {\em instantaneous} rest frame, ${{v}_0}=0$, actually in this case indicates that the convective flow of fields along with the charge that is momentarily stationary, is zero.

Using the vector identity ${\bf v}={\bf n}({\bf v}.{\bf n}) - {\bf n}\times\{{\bf n}\times{\bf v}\}$, we can write the velocity field (first term within the square brackets in Eq.~(\ref{eq:38aa})), in terms of  radial (along {\bf n}) and transverse components (perpendicular to {\bf n}) \cite{17,18},
and then making use of Eq.~(\ref{eq:38c}), we can write the electric field of a uniformly accelerated charge as a sum of a radial term (along {\bf n}) 
\begin{eqnarray}
\label{eq:38aaa1}
{\bf E}_{\rm n}=\left[\frac{e{\bf n}}{\gamma ^2 r^2(1-{\bf n}\cdot{\bf v}/c)^2}\right]_{t_{\rm r}}=e\left[\delta^{2}\frac{{\bf n}}{r^2}\right]_{t_{\rm r}}\:,
\end{eqnarray}
and a transverse term (perpendicular to {\bf n})
\begin{eqnarray}
\label{eq:38aaa}
{\bf E}_{\rm T}&=&\frac{e\gamma_0{v}_0}{c}\left[\frac{{\bf n}\times({\bf n}\times \hat{\bf z})}{\gamma ^{3}r^{2}(1-{\bf n}\cdot{\bf v}/c)^{3}} \right]_{t_{\rm r}}\nonumber\\
&=&\frac{e\gamma_0{v}_0}{c}\left[\delta^{3}\frac{{\bf n}\times({\bf n}\times \hat{\bf z})}{r^{2}} \right]_{t_{\rm r}}\:.
\end{eqnarray}
The transverse term in the electric field, at any space-time event, apart from the relativistic beaming factor $\delta^{3}$ like in the case of the magnetic field, is proportional to the value of the instantaneous relativistic velocity (${\gamma_0{v}_0}$) of the charge, and falls as square of distance ($\propto 1/r^2$), even in the far-off regions.  The transverse electric field of the uniformly accelerated charge is thus zero {\em everywhere} in the instantaneous rest frame, where ${v}_0=0$, consistent with the absence of a magnetic field in this case. 

It cannot happen by a mere chance coincidence that at {\em all space-time} events, the  transverse components of the electromagnetic fields (Eqs.~ (\ref{eq:38bbb}) and (\ref{eq:38aaa})),  for a uniformly accelerated charge, are proportional to the instantaneous velocity of the charge at that time, with the fields falling as square of distance. Thus in the electromagnetic fields of a uniformly accelerated charge there is no term $\propto \dot{\bf v}/r$, that is usually defined as radiation fields.
Actually, the acceleration fields, which are transverse in nature, get cancelled neatly, by a term in the transverse velocity fields, for all $r$, in the case of a uniformly accelerated charge.

The `real-time' field expressions (Eq.~(\ref{eq:38ab1})) are equivalent to the field expressions in terms of retarded-time quantities, and can be derived from  Eqs.~(\ref{eq:38bbb}), (\ref{eq:38aaa1}) and (\ref{eq:38aaa}),  by algebraic transformations \cite{88}.

Equations~(\ref{eq:38bbb}), (\ref{eq:38aaa1}) and (\ref{eq:38aaa}) also describe the electromagnetic field components of a charge moving along $\hat{\bf z}$ with a {\em uniform velocity} $v={v}_0$ (and $\gamma=\gamma_0$), in terms of the time-retarded quantities.

\section{Energy in the electromagnetic fields of a uniformly accelerated charge}
One could pose a legitimate question: Where does the electromagnetic power
represented by acceleration fields go in the case of a uniformly accelerated charge? 

Actually a moving charge has more energy in its electromagnetic fields than a stationary charge. When the velocity of a charge changes due to acceleration, there is a change in its self-field energy. It so happens that at any instant $t$, the contribution of the acceleration fields to the total field energy of a uniformly accelerated charge
makes its self-field energy exactly equal to the value expected in fields because of its instantaneous velocity $v_0$ at that instant.
This could happen because the information about both the velocity and acceleration
of the charge at the retarded time are present in the field expressions (Eq.~(\ref{eq:38aa})),
and as long as the acceleration does not change from its value at the retarded time (a uniform acceleration!), no mismatch in the field energy takes place. 

To comprehend it better, we consider the situation for a retarded time, ${t_{\rm r}}=0$, when the charge was instantaneously at rest (i.e. when $[v]_{t_{\rm r}}=0$) at $z_0$. Actually one can take any other time ${t_{\rm r}} < 0$ or ${t_{\rm r}} > 0$ as the starting point for our discussion, but choosing ${t_{\rm r}}=0$ makes it convenient from computation point of view, as the relativistic beaming factor $[\delta]_{t_{\rm r}}=1$ at ${t_{\rm r}}=0$. In any case, one can always transform to an appropriate inertial frame.
Now, from Eqs.~(\ref{eq:38bbb}), (\ref{eq:38aaa1}) and (\ref{eq:38aaa}), using $[\delta]_{t_{\rm r}}=1$, we can easily calculate the energy in fields of the same uniformly accelerated charge at a later time $t$, when it is moving with a velocity $\gamma_0 v_0 = at$. 

From the formula for the electromagnetic field energy  
\begin{eqnarray}
\label{eq:89.4k}
{\cal E}= \int_{\rm V} \frac{E^{2}+B^{2}}{8\pi}\:{\rm d}V \;,
\end{eqnarray}
the field energy of the uniformly accelerated charge, in a spherical shell ${\Sigma}$ of radius $r$ and thickness ${\rm d}r$ is
\begin{eqnarray}
\label{eq:89.4l}
{\rm d}{\cal E}=\frac{e^2}{8\pi}\int_{\rm o}^{\pi} 2\pi\:{\rm d}\theta\: \left(\sin\theta+\frac{2\gamma_{\rm o}^2{v}_{\rm o}^2}{c^2} \sin^3\theta\right)\frac{{\rm d}r}{r^{2}} \;.
\end{eqnarray}
The electromagnetic field energy in the shell, ${\Sigma}$ turns out to be 
\begin{eqnarray}
\label{eq:89.4m}
{\rm d}{\cal E} =\frac{e^{2}}{2}\left(1+\frac{4\gamma_0^2 v_0^{2}}{3c^2}\right)\,
\frac{{\rm d}r}{r^{2}} \;.
\end{eqnarray}
The first term in parenthesis on the right hand side of Eq.~(\ref{eq:89.4l}) or (\ref{eq:89.4m}) is the energy in the radial field, which is the same as the energy in Coulomb fields of a stationary charge, while the second term, $\propto(\gamma_{0} v_{0})^{2}$, in these equations represents the energy in the transverse fields of the charge, because of its motion. 
Thus the energy in fields of a charge moving with a velocity $v_0$ is higher than that of the stationary ($v_0=0$) charge by an amount $\propto(\gamma_{0} v_{0})^{2}$. 

Now from where does this energy come from? Obviously from the acceleration fields. In fact this extra energy in the shell of radius $r$ and thickness ${\rm d}r$ is 
\begin{equation}
\label{eq:38caa}
{\rm d}{\cal E} =\frac{2e^2}{3c^2}\frac{(\gamma_0 v_0 )^{2}}{r^2}{{\rm d}r}\;,
\end{equation}

In the same way employing the formula for the Poynting vector  
\begin{equation}
\label{eq:8}
{\cal S}= \frac{c}{4\pi}{\bf E}\times {\bf B}\:,
\end{equation}
the electromagnetic power passing through the shell  ${\Sigma}$ is, 
\begin{equation}
\label{eq:38ca}
P= \oint_{\Sigma}{{\rm d}\Sigma}\:({\bf n} \cdot {\cal S})
=\frac{e^2\gamma_{\rm o}^2{v}_{\rm o}^2}{2r^2 c}\int_{\rm o}^{\pi} {\rm d}\theta\: \sin^3\theta
=\frac{2e^{2}\gamma_{\rm o}^2{v}_{\rm o}^{2}}{3r^2c}  \:.
\end{equation}
The power passing through the spherical surface depends on its radius as $1/r^2$.

Thus the Poynting flow (Eq.~(\ref{eq:38ca})), which usually represents Larmor's radiation formula  (with $\gamma_{0} v_0= ar/c$), seems to supply in this case, during a time interval ${\rm d}t$, the exact amount of field energy, in a shell of  thickness ${\rm d}r=c{\rm d}t$ (Eq.~(\ref{eq:38caa})), for all $r$, pertaining to the instantaneous velocity, $v_0$, of the charge. 
The acceleration fields, in this way, continuously keep updating the self-fields, for the latter to remain in conformity with the changing velocity of the charge. 
Apart from this, there is {\em no other field energy}, {\em emitted away} as radiation from a uniformly accelerated charge.

It is of course possible to calculate the electromagnetic field energy in the case of a charge moving with a uniform velocity $v_0$ or even of a charge moving with a uniform acceleration, for the case with ${t_{\rm r}} \ne 0$ as well. In either case, for instance, we obtain the same analytical expression Eq.~(\ref{eq:89.4m}), where $\gamma_0 v_0$ correspond to the instantaneous motion in the uniform acceleration case \cite{18}. Thus it is clear that the acceleration fields in the case of a  uniformly accelerated charge add (or even subtract during  a deceleration) just sufficient energy in fields so as to make the total field energy equal to that required at any instant according  to the instantaneous velocity of the accelerating charge at that moment.

In the same way, one can calculate the Poynting flux in the case of a charge moving with a uniform velocity $v_0$ or even of a charge moving with a uniform acceleration with an instantaneous velocity $v_0$ starting from ${t_{\rm r}} \ne 0$ \cite{18}. Usually, the Poynting flux through a spherical surface of radius $r$, in unit time $t_{\rm r}$ of the charge, gives the radiated power, ${2e^{2}a^2}/{3c^3}$, which is independent of distance $r$ from the charge. 
However, in the case of a uniformly accelerated charge, or even of a charge moving with a uniform velocity, 
we see that the Poynting flux falls rapidly with distance (${\bf S} \propto 1/r^2$) with no term proportional to $a^{2}$ and independent of $r$, implying thereby, no radiation from a uniformly accelerated charge.

\section{The relation between the charge movement and the Poynting flow}
\begin{figure}[t]
\begin{center}
\includegraphics[width=\columnwidth]{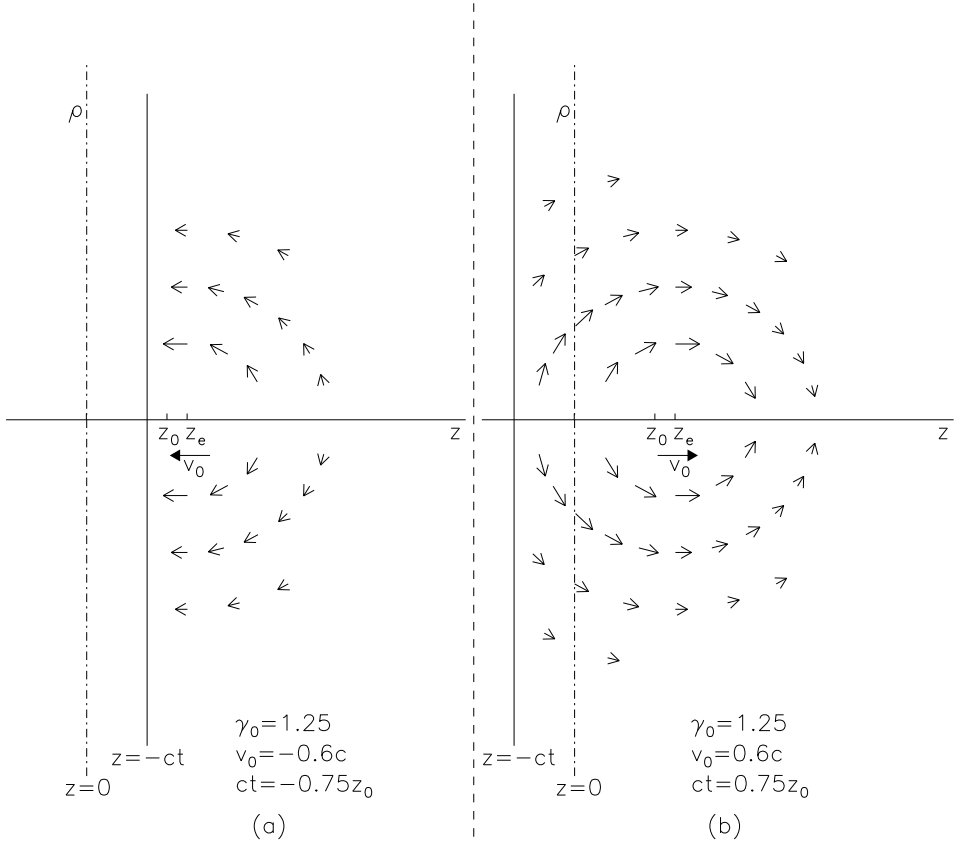}
\caption{The transverse Poynting vector component, $S_\Theta$, around a charge moving with a uniform proper acceleration along the $z$-axis, for the two cases shown in Fig.~1 (a) and (b). Arrows show Poynting vector directions at different distances from the charge. 
In this schematic diagram, the length of an arrow is not directly proportional to the magnitude of the corresponding Poynting vector, the plot merely indicates the trend qualitatively, with  the magnitude of the Poynting vector in successive concentric outer circle falling by a factor of $\sim 10$. The overall Poynting flow is along the instantaneous velocity of the charge.}
\end{center}
\end{figure}
\begin{figure}[t]
\begin{center}
\includegraphics[width=\columnwidth]{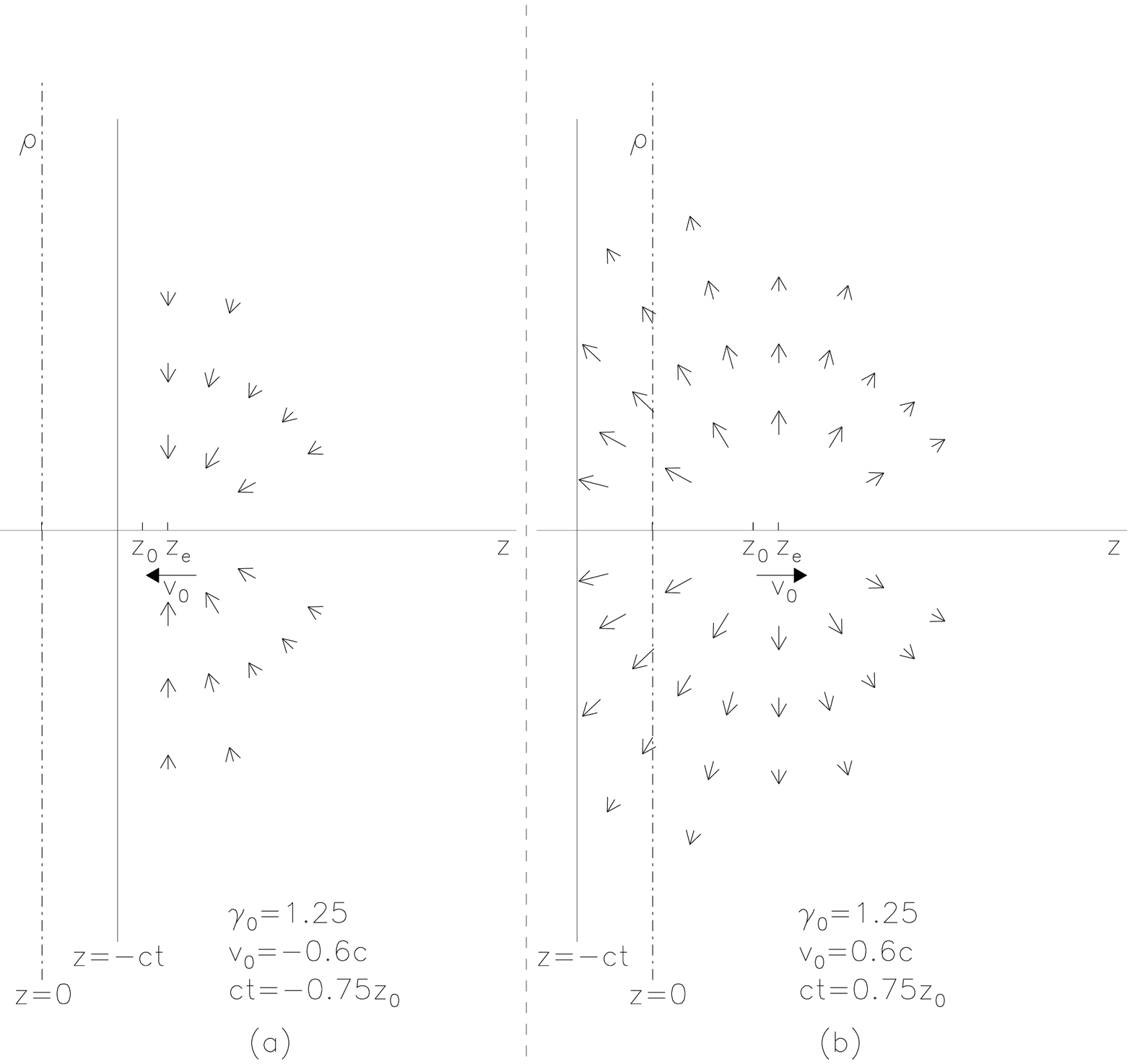}
\caption{The radial Poynting vector component, $S_{\rm R}$, for a charge moving with a uniform proper acceleration along the $z$-axis, for the two cases shown in Fig.~1 (a) and (b). Arrows show Poynting vector directions at different distances from the charge.  
In this schematic diagram, the length of an arrow is not directly proportional to the magnitude of the corresponding Poynting vector, the plot merely indicates the trend qualitatively, with the magnitude of the Poynting vector in successive concentric outer circle falling by a factor of $\sim 10$. When the charge is decelerating, as in (a), the Poynting vector is everywhere radially inward toward $z_{\rm e}$, the present charge position, while in (b), where the charge is accelerating, the Poynting vector is everywhere radially outward from  $z_{\rm e}$.}
\end{center}
\end{figure}
As the uniformly accelerated charge at time $t$ is located at $z_{\rm e}=[{z_0^2+c^2t^2}]^{1/2}$, we can express Eq.~(\ref{eq:38ab1}) in terms of $z_{\rm e}$, the present position of the charge, as 
\begin{eqnarray}
\label{eq:89ab1}
E_{z}&=&-4ez_0^{2}(z_{\rm e}^2+\rho^{2}-z^{2})/\xi^{3}\nonumber\\
E_{\rho}&=&8ez_0^{2}\rho z/\xi^{3}\nonumber\\
B_{\phi}&=&8ez_0^{2}\rho ct/\xi^{3}\;,
\end{eqnarray}
with $\xi=[(z_{\rm e}^2-\rho^{2}-z^{2})^{2}+4z_0^{2}\rho^{2}]^{1/2}$.

The cylindrical symmetric expressions (Eq.~(\ref{eq:89ab1})) for electromagnetic fields, can be expressed in polar coordinates ($R,\Theta,\Phi$), centered on the instantaneous charge position,  $z_{\rm e}$ \cite{18}. Substituting $z=z_{\rm e}+R \cos\Theta$, $\rho=R \sin\Theta$ and $v_0/c=ct/z_{\rm e}$ in Eq.~(\ref{eq:89ab1}), and after some algebraic manipulations, we get
\begin{eqnarray}
\label{eq:38ab3}
E_{\rm R} &=&\frac{e(1+\eta \cos\Theta)}{R^{2}\gamma^{2}
\{(1+\eta\cos\Theta)^{2}+(\eta^{2}-(v_0/c)^{2})\sin^{2}\Theta\}^{3/2}}
\nonumber\\
E_{\Theta}&=&\frac{e\eta \sin\Theta}{R^{2}\gamma^{2}
\{(1+\eta\cos\Theta^{2}+(\eta^{2}-(v_0/c)^{2})\sin^{2}\Theta\}^{3/2}}
\nonumber\\
B_{\Phi}&=&\frac{e(v_0/c) \sin\Theta}{R^{2}\gamma^{2}
\{(1+\eta\cos\Theta^{2}+(\eta^{2}-(v_0/c)^{2})\sin^{2}\Theta\}^{3/2}}\;,\nonumber\\
\end{eqnarray}
where $\eta=a R/2\gamma_0 c^{2}=R/2\gamma_0 z_0=R/2z_{\rm e}$. The remaining field components are all zero.

The Poynting vector can be decomposed into transverse (along ${\Theta}$) and radial (along ${R}$) components
\begin{equation}
\label{eq:38ab5a}
S_{\Theta}=\frac{-c}{4\pi}E_{\rm R}B_{\Phi}
\end{equation}
\begin{equation}
\label{eq:38ab5b}
S_{\rm R}=\frac{c}{4\pi}E_{\Theta}B_{\Phi}
\end{equation}

Figure~3 shows the transverse Poynting vector component, $S_\Theta$ about the uniformly  accelerated  charge, presently at $z_{\rm e}$. Arrows shows the transverse Poynting vector directions at different distances from the charge, at times $t=\pm 0.75 z_0/c$ when the charge was moving, respectively, with velocity $v=\pm 0.6c$. The charge at both events has a corresponding Lorentz factor, $\gamma=1.25$. The overall Poynting flow in either case is along the instantaneous velocity of the charge. As we will discuss later in Section 6, the Poynting flows in Fig.~3, represent the convective flow of the self-field energy along with the moving charge. The $z=-ct$ plane, in each case, denotes the causality limit of fields.

Figure~4 shows the radial Poynting vector component, $S_{\rm R}$, for times $t=\pm 0.75 z_0/c$, the same as in Fig.~3, with arrows showing radial Poynting vector directions at different distances from the charge.  At the time $t=-0.75 z_0/c$, when the charge was decelerating, the Poynting vector is everywhere radially inward toward  $z_{\rm e}$, the present charge position, while at time $t=0.75 z_0/c$ when the charge is accelerating, the Poynting vector is  everywhere radially outward from  $z_{\rm e}$.

From a comparison of the radial Poynting vectors from Fig.~4, along with Eqs.~(\ref{eq:38ab3})) and (\ref{eq:38ab5b})), it is evident that along with any change taking place in the velocity of the charge, the change in its self-field energy at time $t$ is equal and opposite to that occurring at $-t$. While at time $t>0$ the fields grow, with energy in fields increasing along with the  increasing velocity (acceleration) of the charge, at time $t<0$ energy in fields diminishes proportionally as the fields weaken with a decreasing velocity (deceleration) of the charge, and with a nil rate of change in the field strength as well as in the field energy at $t=0$. 

\section{The change in electric field configuration as velocity of the charge changes due to its uniform acceleration}
Usually the picture for a radiating charge one has in mind is that the radiation (electromagnetic wave) moves toward infinity leaving the source (charge) behind, close to where it was at the time of emission, e.g., the radiating charges within an antenna. That indeed would be the case if it is a bound motion which necessarily will have velocity over a certain, sufficiently large, period of time getting averaged to zero. In such cases one does not bother much about the velocity fields as well as about the energy in the self-fields of a charge. However the picture may be quite different in the case of an unbound motion. A charge moving with a uniform velocity is the simplest example of an unbound motion, where the charge may even move to infinity, from its initial position, during a large enough time interval. There is a finite radial Poynting flow ($\propto v_0^2/r^2$) at a distance $r$ from the time-retarded position of the charge, having an instantaneous velocity $v_0$, which of course is uniform in this case. However, there is no radiation when seen with respect to the instantaneous position of the charge. Quite similar is the case of a uniformly accelerated charge, which is also an unbounded motion. There is also a finite Poynting flow ($\propto v_0^2/r^2$) at a distance $r$ (Eq.~(\ref{eq:38ca})) from the time-retarded position of the charge, moving with an  instantaneous velocity $v_0$. Though in this case the instantaneous value of velocity is different at different instants because of the uniform acceleration, yet in this case too no radiation takes place \cite{FE95}. If there is an outward radial Poynting flow during the acceleration phase, there is an equal {\em inward} Poynting flow during the deceleration phase. Moreover, as we will show, in this case the charge may not be deserted far behind when the fields move toward infinity. 

To examine the field configuration at large $r=ct$, vis-\`a-vis the location of the uniformly accelerated charge at that time, we need to consider that for large $t$, the Lorentz factor $\gamma_0$ of the charge will be large too. Now, the fields (Eq.~(\ref{eq:89ab1})) will be stronger when $\xi$ is smaller, which would happen, for any given $\rho$, in the vicinity of $z_{\rm e}^2-z^{2}=\rho^{2}$, where $\xi=2 z_0 \rho$ is the lowest value. Let us shift the origin to the `present' position, $z_{\rm e}=\gamma_0 z_0$, of the charge and write $z=z_{\rm e}+\Delta z$. Then $\xi$ will be small and the fields appreciable, for a large $\gamma_0$, only when  
\begin{eqnarray}
\label{eq:89ab5}
|(2z_{\rm e}\Delta z)|\stackrel{<}{\sim} \rho^{2}+(\Delta z)^2\,,
\end{eqnarray}
or, to a first order in $\Delta z$, we have $|\Delta z|\stackrel{<}{\sim} \rho^{2}/2z_{\rm e}=\rho^{2}/2\gamma_0 z_0$, implying $|z_{\rm e}^2+\rho^{2}-z^{2}|\stackrel{<}{\sim} 2\rho^{2}$.  

In that region, from Eq.~(\ref{eq:89ab1}), the ratio of field components, for a large $\gamma_0$, will be
\begin{eqnarray}
\label{eq:89ab6}
\left|\frac{E_{\rho}}{E_{z}}\right| = \left|\frac{2\rho z}{z_{\rm e}^2+\rho^{2}-z^{2}}\right| \sim\frac{2\rho z_{\rm e}}{2\rho^{2}}=\frac{ z_0 \gamma_0}{\rho}\,.
\end{eqnarray}

Thus the electric field, appreciable mostly in a narrow region $|\Delta z|\stackrel{<}{\sim} \rho^{2}/2\gamma_0 z_0$, around the charge, is relatively much stronger in a direction perpendicular to z-axis, the direction of motion of the charge. The field, being all around the instantaneous location of the charge, appears to be like that of a charge moving with a relativistic {\em uniform} velocity $v_0$ along the $z$-axis, with the corresponding Lorentz factor $\gamma_0\gg 1$, and where the field is appreciable around the instantaneous position of the charge only in a narrow $z$ range, $\propto 1/\gamma_0$, with the  $E_{\rho}$ component being stronger than the $E_{z}$ component by a factor $\gamma_0$. 

When expressed in terms of the polar coordinates ($R,\Theta)$ of Eq.~(\ref{eq:38ab3}), we can rewrite the condition in Eq.~(\ref{eq:89ab5}) to state that for $\gamma_0 \gg 1$, the fields would be rather weak except in a short range, $\Delta z=R\cos\Theta \stackrel{<}{\sim} R^{2}/(2\gamma_0 z_0)$, or $\cos\Theta \stackrel{<}{\sim} \eta$. 
Moreover, in the finite region around the charge, $\gamma_0 \gg 1$ implies $\eta \ll 1$. 

With $\gamma_0$ increasing, the fields tend to be those of a charge moving with a {\em uniform velocity} $v_0$ can be also understood in the following way. First thing to note is that with increasing time $t$, as the velocity of the charge, due to a constant acceleration, approaches $c$ and the Lorentz factor becomes very large ($\gamma_0 \rightarrow \infty$), the factor $\eta\propto 1/\gamma_0\rightarrow 0$. Then in regions around the charge, where the fields, falling as $1/R^2$, may only be appreciable for finite, smaller $R$, the transverse component $E_{\Theta}$ becomes increasingly weak as compared to the radial component $E_{\rm R}$ as
\begin{eqnarray}
\label{eq:38ab3.1}
\frac{E_{\Theta}}{E_{\rm R}}=\frac{\eta \sin\Theta}{(1+\eta \cos\Theta)}\,,
\end{eqnarray}
and the electromagnetic fields of the uniformly accelerated charge in Eq.~(\ref{eq:38ab3}) will be reducing to those of a charge moving with a uniform velocity $v_0$, i.e., the present velocity of the uniformly accelerated charge, given by  \cite{1,2,25,PU85}
\begin{eqnarray}
\nonumber
{\bf E}=\frac{e\hat{\bf R}}{R^{2}\gamma_0^{2}[1-(v_0/c)^{2}\sin^{2}\Theta]^{3/2}}\\
\label{eq:38ab4}
=\frac{e\gamma_{\rm 0}{\bf R}}{[\gamma_{\rm 0}^2\Delta z^{2}+\rho^2]^{3/2}}\,,
\end{eqnarray}
with the magnetic field, ${\bf B}= {\bf v}_0\times {\bf E}/c$.

In Eq.~(\ref{eq:38ab4}), for  any  given $\rho$, the denominator will be small and the field appreciable for a large values of $\gamma_0$, only when  $|\Delta z|\stackrel{<}{\sim} \rho/\gamma_0$ and in that region, ratio of the field components, for a large $\gamma_0$, will be
\begin{eqnarray}
\label{eq:89ab7}
\left|\frac{E_{\rho}}{E_{z}}\right| = \left|\frac{\rho}{\Delta z}\right| \sim \gamma_0\,.
\end{eqnarray}
  
Thus the electric field is relatively stronger by a factor $\gamma_0$ in a direction perpendicular to $z$-axis, the direction of motion of the charge, and is confined mostly to a narrow region $\Delta z\sim \rho/\gamma_0$ around the charge. Therefore, for a large $\gamma$, the field lines becoming oriented perpendicular to the direction of motion \cite{1,2,25}, concentrated mostly within a small angle, $\Delta z/\rho \sim 1/\gamma$, with respect to a plane normal to the direction of motion and passing through the instantaneous charge position \cite{88}. 
Thus apart from a scaling factor of $\sim \rho/z_0=a \rho/c^2$, the field configuration of a uniformly accelerated charge (Eq.~(\ref{eq:89ab6})), when $\gamma_0$ is large, is quite similar to that of a a uniformly moving charge having the same $\gamma_0$ factor (Eq.~(\ref{eq:89ab7})).

Of course, in the close vicinity of the charge, where $R\rightarrow 0$, implying $\eta\rightarrow 0$, the field lines from the charge in all cases will emerge like that from a charge with uniform velocity $v_0$.
the fields reduce to that of a charge moving with a uniform velocity ${\bf v}_0$, with the electric field everywhere in a radial direction from the present position of the charge,  falling as $1/R^2$. But with $\gamma_0$ becoming larger, the electric field lines, like those of a uniformly moving charge, increasing get oriented perpendicular to the direction of motion of the charge

Like in the case of a uniformly moving charge, where the fields determined from the time-retarded positions of the charge, are always radial from the `instantaneous' position of the charge, in the case of a uniformly accelerated charge, the situation is not very different. Here too we find the electric field initially starting in radial directions from the `instantaneous' position of the charge, however, there is some bending in the field lines, at larger distances, depending upon the ratio of the instantaneous velocity of the charge and the magnitude of the constant acceleration. However, with increasing $\gamma_0$, the bending reduces, and the field becomes more like that of a uniformly moving charge. Figure~5 shows the electric field distribution of a charge uniformly accelerated and moving presently with a velocity $v_0=-0.995c$, as well as of a charge moving with a present velocity $v_0=0.995c$, both corresponding to $\gamma_0=10$. A comparison with Fig.~1 shows the change in the field distributions, which with much larger $\gamma_0$ become increasingly like that of a charge moving with a uniform velocity with the same large $\gamma_0$, as shown in Fig.~5(c). 
\begin{figure*}[t]
\begin{center}
\includegraphics[width=12.cm]{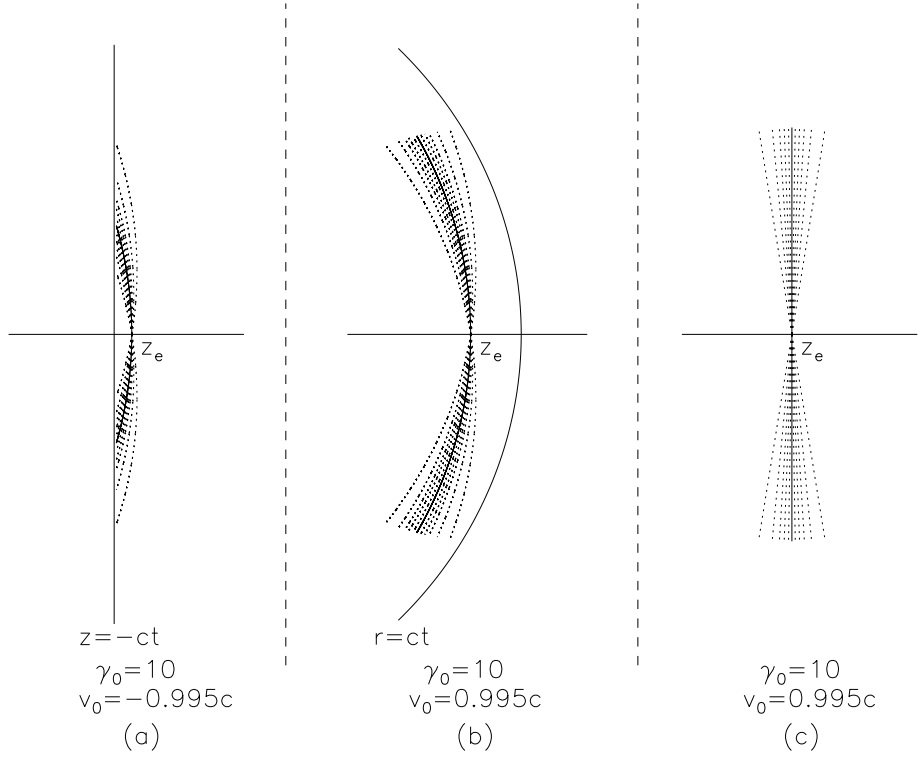}
\caption{The electric field distribution of a charge (a) uniformly accelerated and moving presently with a velocity $v_0=-0.995c$, corresponding to $\gamma_0=10$, with the $z=-ct$ plane showing relatively the causality limit of fields (b) uniformly accelerated and moving presently with a velocity $v_0=0.995c$, corresponding to $\gamma_0=10$, with $r=ct$ showing the relative position of the leading spherical light-front (c) moving with a uniform velocity $v=0.995c$, corresponding to $\gamma=10$. In each case, $z_{\rm e}$ marks the instantaneous position of the charge.}
\end{center}
\end{figure*}

\begin{figure*}[t]
\begin{center}
\includegraphics[width=11.cm]{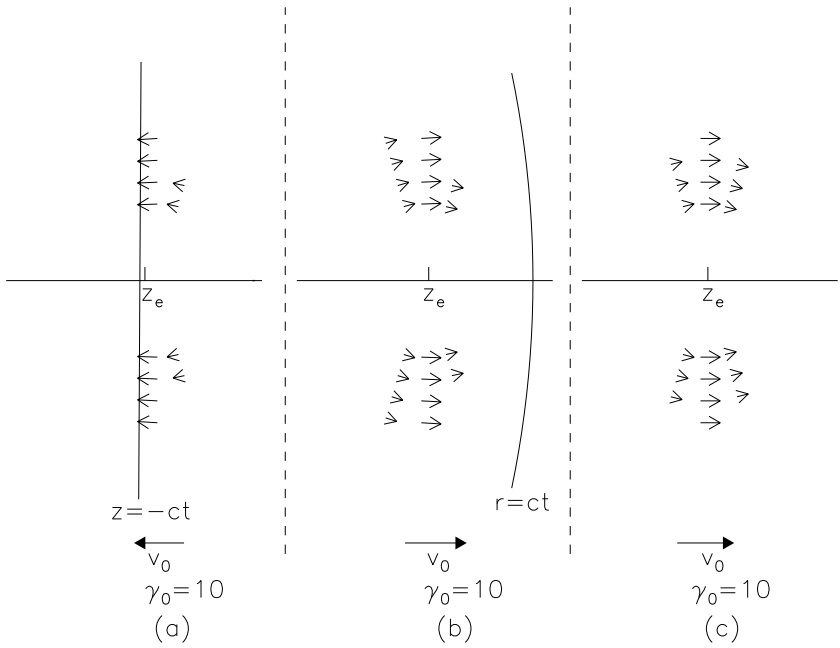}
\caption{The transverse Poynting vector component, $S_\Theta$, around a charge moving with a uniform proper acceleration along the $z$-axis. Arrows show Poynting vector directions at different distances from the charge. (a) The charge is moving with an instantaneous velocity $v_0=-0.995c$ and is getting decelerated, with the $z=-ct$ plane showing relatively the causality limit of fields. (b) The charge is moving with an instantaneous velocity $v_0=0.995c$, and is getting accelerated, with the leading spherical `light-front' at  $r=ct$. (c)  The charge is moving with a uniform velocity $v=0.995c$. In all three cases the corresponding Lorentz factor, $\gamma$, is 10.  The overall Poynting flow is along the instantaneous velocity vector of the charge in all cases.  In this schematic diagram, the length of an arrow is not a direct indicator of the magnitude of the corresponding Poynting vector, the plot merely shows the trend qualitatively. In fact, the magnitude of the Poynting vector, represented by larger arrows, is maximum at the vertical plane, i.e., at the plane normal to the direction of motion, passing through the charge at $z_{\rm e}$,  and drops  rapidly off the plane. At the positions of smaller arrows seen in the figure, the magnitude of the Poynting vector in all three cases falls as much as by a factor of  $\sim 10^4$ compared to that at the vertical plane through $z_{\rm e}$.}
\end{center}
\end{figure*}

\begin{figure}[t]
\begin{center}
\includegraphics[width=7cm]{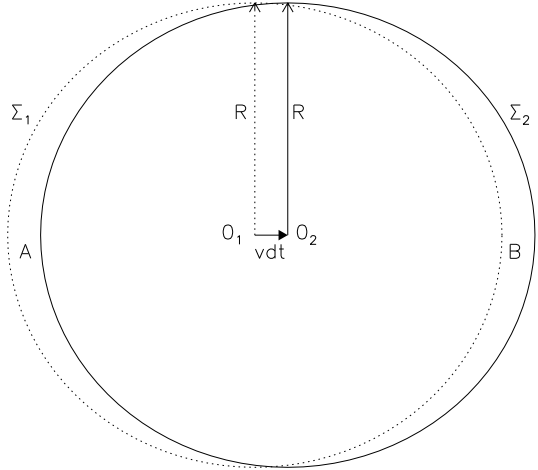}
\caption{As a charge moves with a velocity $v$ from its position $O_1$ to $O_2$, its self-fields also move with it. If we consider two spherical volumes, $\Sigma_1$ and $\Sigma_2$, each of radius $R$ around the two charge positions, the field energy in the region of intersection $B$ between the two spheres increases at the cost of the field energy in the region $A$, where it reduces. The Poynting flows in Figs. 3 as well as 6, represent the convective flow of the self-field energy along with the moving charge.}
\end{center}
\end{figure}
\begin{figure}[t]
\begin{center}
\includegraphics[width=7cm]{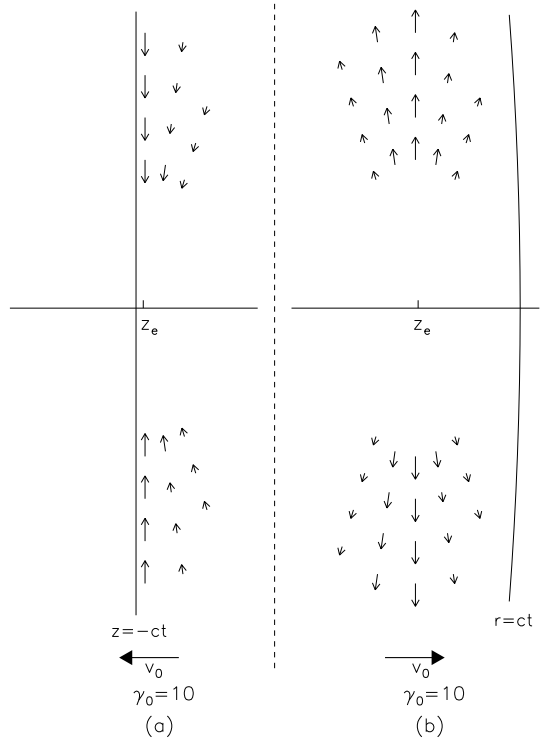}
\caption{The radial Poynting vector component, $S_{\rm R}$, for a charge moving with a uniform proper acceleration, and is presently at $z_{\rm e}$. Arrows show Poynting vector directions at different distances from the charge. (a) The charge is moving with an instantaneous velocity $v_0=-0.995c$, and actually getting decelerated with the Poynting flow being {\em everywhere inward} toward the charge at $z_{\rm e}$, with the $z=-ct$ plane showing relatively the causality limit of fields. (b) The charge is moving with an instantaneous velocity $v_0=0.995c$, corresponding to $\gamma_0=10$, the Poynting flow being outward from the accelerating charge at $z_{\rm e}$. The leading spherical light-front $r=ct$ is shown with respect to $z_{\rm e}$.   In this schematic diagram, the length of an arrow is not directly proportional to the magnitude of the corresponding Poynting vector, the plot merely indicates the trend qualitatively. In fact, the magnitude of the Poynting vector, represented by larger arrows, is maximum at the plane normal to the direction of motion, passing through the charge at $z_{\rm o}$,  and drops  rapidly off the plane. At the positions of smallest arrows seen in the figure, the magnitude of the Poynting vector in both cases falls as much as by a factor of  $\sim 10^4$.}
\end{center}
\end{figure}

\begin{figure*}[t]
\begin{center}
\includegraphics[width=13.5cm]{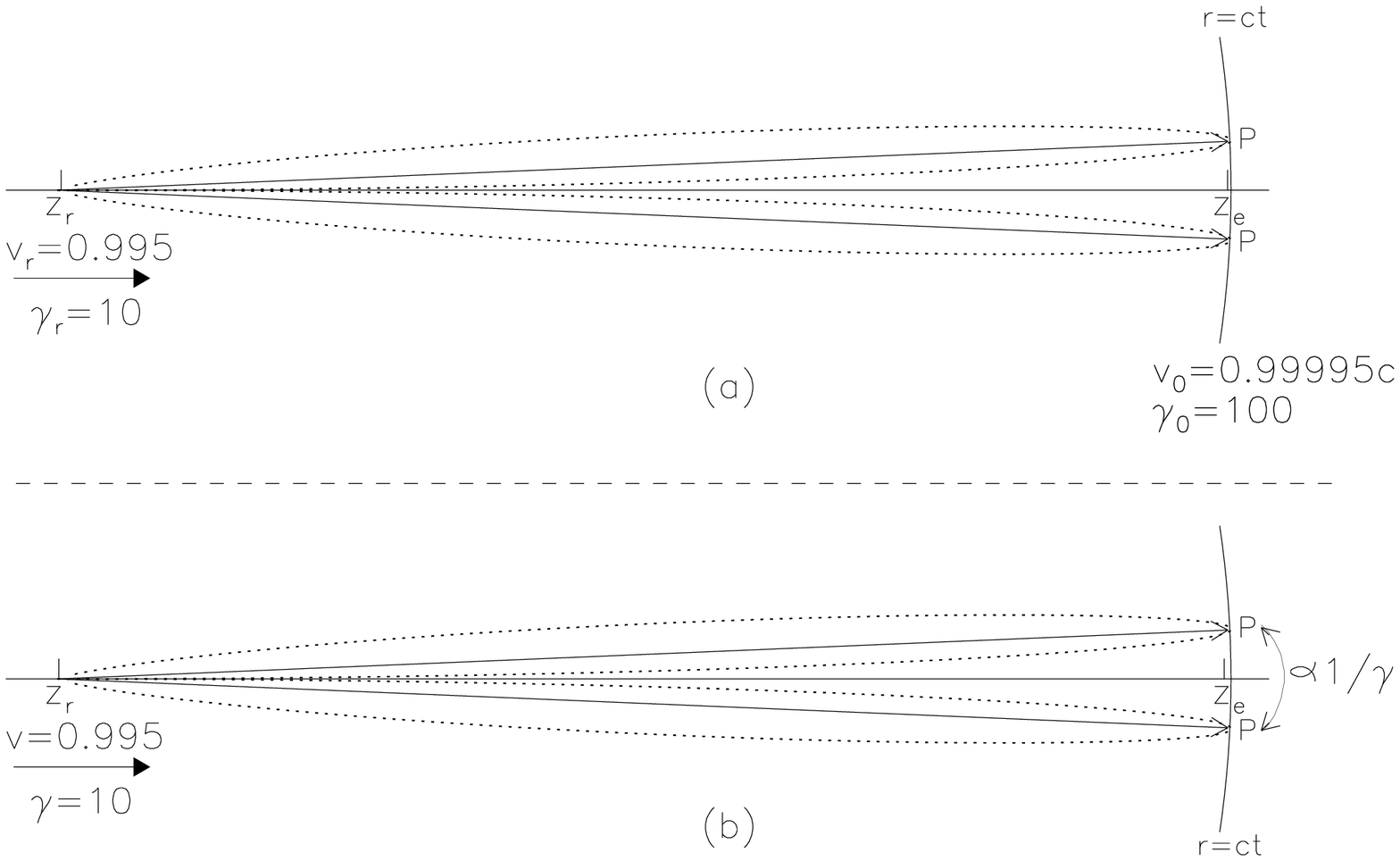}
\caption{Angular distribution of the electric field strength with respect to the time-retarded position $z_{\rm r}$ of the charge, moving along the $z$-axis with a velocity $v=0.995$ and the corresponding Lorentz factor $\gamma=10$. The maxima of the field strength is along the direction  $z_{\rm r}P$, which is at an angle $\theta = 1/\sqrt 5\gamma$. Thus the field distribution mostly lies with a cone of angle $\theta \sim 1/\gamma$. By the time the fields from the retarded position $z_{\rm r}$ reach at the field point P, the charge moves to $z_{\rm e}$. The maxima of the electric field with respect to instantaneous position $z_{\rm e}$ of the charge seems to be in a plane perpendicular to the $z$-axis. The upper panel (a) is for the uniformly accelerated charge, while the lower panel (b) is for a charge moving with a uniform velocity. There is hardly any difference in the two panels, except that in the case (a) the instantaneous charge position is slightly nearer to the leading spherical light-front where because of the acceleration the  velocity has now, say, become $v_0=0.99995$, with $\gamma_0=100$, whereas in case (b) the velocity remains constant at $v=0.995$.}
\end{center}
\end{figure*}

Figure~(6) shows the transverse Poynting vector component, $S_\Theta$, for the uniformly accelerated charge moving with instantaneous velocities $v_0=\pm 0.995c$, with respect to the instantaneous position, $z_{\rm e}$, of the charge in each case. If we compare it with the Poynting flow about a charge moving with a uniform velocity $v=0.995c$ (Fig.~(6c)), we find the Poynting flow to be similar in all cases, with the overall Poynting flow being along the direction of motion of the charge in each case.

In a uniform velocity case, there is no radiation emitted by the charge, even though the Poynting flow is non-zero. The Poynting flow in Fig. (6c), represents the convective flow of the self-field energy along with the moving charge.
Figure~7 explains it pictorially. As a charge with a uniform velocity moves from its position $O_1$ to $O_2$, its self-fields also move with it, as if `attached' to it. If we consider two spherical volumes, $\Sigma_1$ and $\Sigma_2$, each of radius $R$ around the two charge positions, the field energy in the region of intersection $B$ between the two spheres increases at the cost of the field energy in the region $A$, where it reduces. Likewise, in the uniform  acceleration case too, the fields move along with the charge, as demonstrated by the Poynting flows in Fig. (3) as well as Fig. (6a,b), representing the convective flow of the self-field energy along with the moving charge. 
The Poynting flux inferred by Fulton and Rohrlich \cite{5}, calculated with respect to the time-retarded position of the charge is mainly the convective flow of the self-fields, about the instantaneous 'present' position of the charge, that move along with the charge, like in the case of a charge moving with a uniform velocity. 

The only difference in the case of a uniformly  accelerated charge is that there is a radial Poynting flow from the present position of the charge, outward in the case of acceleration (that is when the velocity of the charge is increasing and thereby the energy in its self-fields is increasing) and an inward flow toward the present position of the charge when it is getting decelerated and the energy in its self-fields is decreasing (Fig.~8). This radial flow maintains the field strength and the consequential self-field energy of a uniformly  accelerated charge in synchronism with its changing velocity due to its constant acceleration.

\begin{figure*}[t]
\begin{center}
\includegraphics[width=13.5cm]{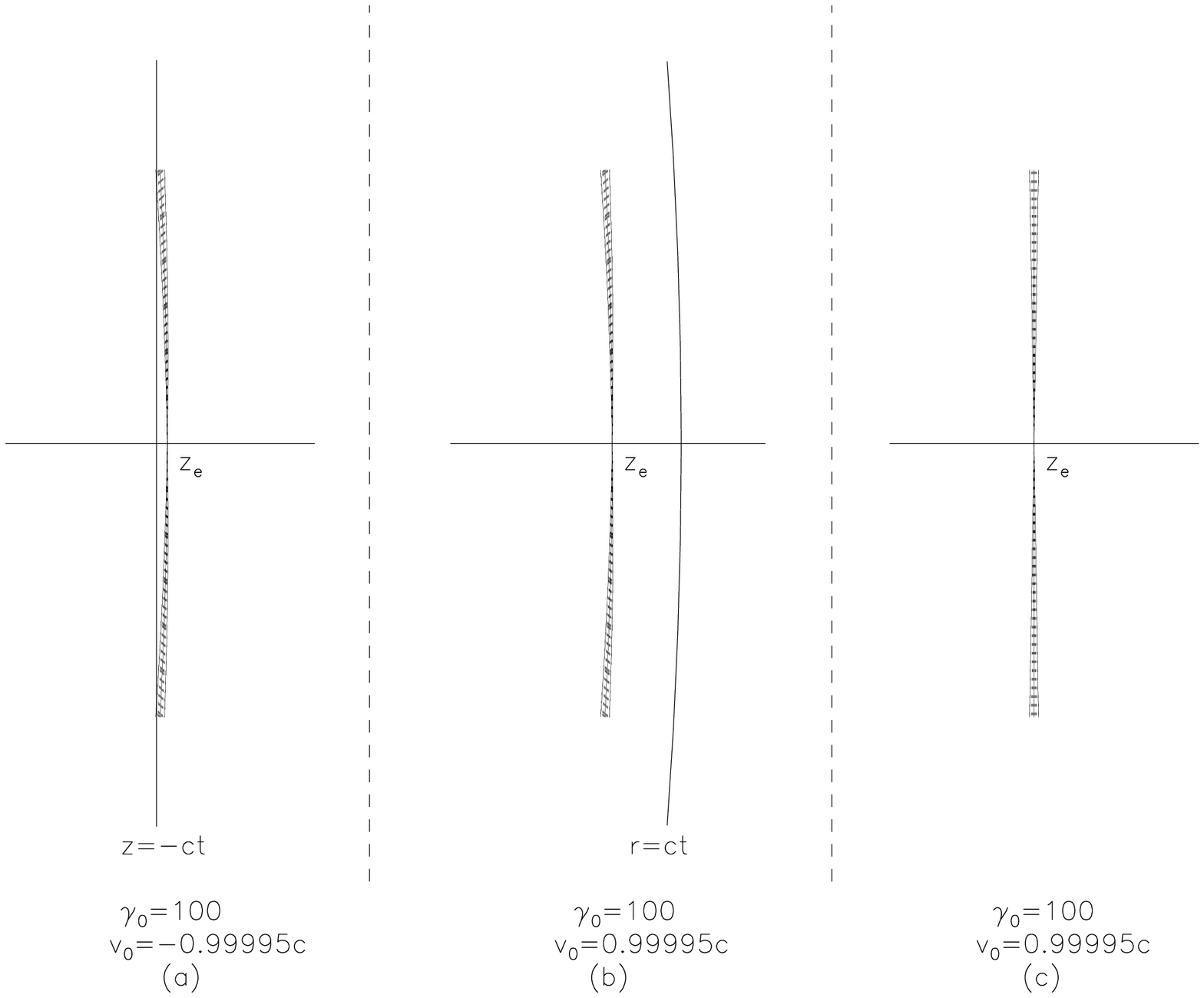}
\caption{The electric field distribution (a) of a uniformly accelerated charge, with an instantaneous velocity $v_0=-0.99995 c$, corresponding to $\gamma_0=100$, with the $z=-ct$ plane showing the causality limit of fields with respect to $z_{\rm e}$, the `present position of the  charge (b) of a uniformly accelerated charge, with an instantaneous velocity $v_0=0.99995 c$, corresponding to $\gamma_0=100$, with $r=ct$ showing the leading spherical light-front with respect to $z_{\rm e}$ (c) of a charge, presently at $z_{\rm e}$, moving with a uniform velocity $v_0=0.99995 c$, corresponding to $\gamma_0=100$. In all three cases, the electric field lines are confined mostly within a small angle $\sim 1/\gamma_0=10^{-2}$, with respect to the electric field lines that begin from the instantaneous charge position, $z_{\rm e}$, in  plane perpendicular to the direction of motion.} 
\end{center}
\end{figure*}

\section{Where lies the charge when the Poynting flow is at infinity?}
In a normal radiation scenario, when the radiation is  examined at a large distance ($r\rightarrow \infty$!), the charge responsible is nowhere in the vicinity. However, for a uniformly accelerated charge, the situation is quite different. At large $t$, speed of the uniformly accelerated charge approaches that of light and it is not very far behind the leading spherical front of the Poynting flux at $r=ct$, advancing with speed $c$. For a large $\gamma$, the charge, moving with a velocity $v\approx c(1-1/2\gamma^2)$, follows behind, merely a distance, $ct/2\gamma^2$, from the leading spherical light-front, 
which due to a large relativistic beaming factor, lies in a narrow cone about the direction of motion. We showed earlier that the electric field configuration resembles closely that of a uniformly moving charge with velocity equal to the instantaneous velocity of the uniformly accelerated charge, with electric field being from the instantaneous charge position almost in a perpendicular direction with respect to the direction of motion. 

In order to understand how fields appear in vertical direction with respect to the instantaneous charge position, we determine the maxima of the angular distribution of field strength with respect to the charge position at the corresponding retarded time.
From Eq.~(\ref{eq:38aaa})), the angular dependence of the transverse electric field component, $E_{\rm T}$, of a charge moving with a velocity $\bf v$, with respect to the time-retarded charge position, is given by 
\begin{eqnarray}
\label{eq:89a}
E_{\rm T}(\theta)\propto\frac{\sin\theta}{(1-v\cos\theta/c)^{3}}\:,
\end{eqnarray}
where $\theta$ is the angle with respect to the direction of motion. Maximizing $E_{\rm T}$ with respect to $\theta$, gives
\begin{eqnarray}
\label{eq:89b}
\cos\theta_{\rm m}=\frac{c}{4v}\left(\sqrt{(1+24(v/c)^2}-1\right)\:,
\end{eqnarray}
from which for $v\rightarrow c$, we get 
\begin{eqnarray}
\label{eq:89c}
\theta_{\rm m}=\frac{1}{\sqrt{5}\gamma}\:,
\end{eqnarray}
The field strength has a maximum at $\theta_{\rm m}=1/\sqrt 5\gamma$, in fact the field distribution is confined mostly in a cone with an opening angle $\theta \sim 1/\gamma$. We have shown  in Fig.~9 the angular distribution of field strength, plotted with respect to the charge position at the corresponding retarded time, where the charge had a relativistic motion $v=0.995c$, with a  corresponding Lorentz factor $\gamma=10$, and for which $\cos\theta_{\rm m}=0.999$. The circle represented by points $P$ on the leading spherical light-front $r=ct$ where the field strength is maximum as a function of $\theta$, lies almost vertically above $z_{\rm e}$, the instantaneous charge position in the case of a charge moving with a uniform velocity $v$ as shown in Fig.~9(b). 

In the case of a charge moving with a uniform acceleration, the angular distribution of field strength plotted with respect to the charge position at the corresponding retarded time is exactly the same, while  $z_{\rm e}$, the instantaneous charge position in this case (Fig.~9(a)), lies even closer to the leading spherical light-front at $r=ct$.  
Fig.~10 shows the field distribution with respect to the instantaneous charge positions in each case. As we had mentioned earlier, the bending in the electric field lines reduces with increasing $\gamma_0$, which is seen from a comparison of Figs.~1, 5 and 10, in the later cases, the field distributions, which with much larger $\gamma_0$, become increasingly like those of a charge moving with a uniform velocity with the same large $\gamma_0$.
As can be seen from Fig.~10(a), at $t<0$, the plane $z=-ct$, which is the boundary of the causality domain of the fields, is almost passing through the instantaneous charge position (cf. Eq.~(\ref{eq:38ab2})). On the other hand, as shown in Fig.~10(b) at $t>0$, while the leading spherical light-front has moved a distance $r=ct$, the charge meanwhile has moved from $z_0$ to $z_{\rm e}$, and lies at a distance from the intersection of  the light-front on the $z$-axis as
\begin{eqnarray}
\label{eq:38ab6}
ct-(z_{\rm e}-z_0)&=&\gamma_0 z_0 v_0/c-z_0(\gamma_0-1)\nonumber\\
&\approx&z_0[1- 1/2\gamma_0]\stackrel{<}{\sim} z_0\,, 
\end{eqnarray}
Thus even when the electromagnetic fields, including the acceleration fields, are at infinite distances ($r\rightarrow \infty$) from the time-retarded position of the charge, they may still be at finite distance from the instantaneous position of the charge, which itself is moving toward infinity due to the uniform acceleration, with the fields appreciable in a region $\Delta z$ around the charge that shrinks as $\propto 1/\gamma_0$, for large $\gamma_0$. 

We saw from Eq.~(\ref{eq:38ca}) that the Poynting flux for  a uniformly accelerated charge is the same as of a uniformly moving charge, and therefore, like in the latter case, the Poynting flow in the case of a uniformly accelerated charge too represents ``convective'' flow of fields due to the ``present velocity'' of the charge. Thus the Poynting flux at large $r$ (infinity!) cannot be termed as ``emitted'' by the charge, instead it belongs to the charge as the ``convective'' flow of its self-fields due to the ``present velocity'' of the charge. 
Additionally, in case of a uniformly accelerated charge a part of the Poynting flux also accounts for the change occurring in the self-field energy of the charge, because of the ever changing velocity of the charge.
After all the charge now moving with $v_0=0.99995$ and $\gamma_0=100$, with self-field energy $\propto v_0^2\gamma_0^2=10^4$ was at $t=0$ with $v_0=0.$ and $\gamma_0=1$ with minimal self-field energy. This growth in self-field energy could come where else from but the acceleration fields, which do not represent a radiation being `emitted away' by the uniformly accelerated charge. Naturally there is no radiation reaction in this case.

\section{Discussion}
In the case of a uniformly accelerated charge, it is almost universally accepted that there is no radiation reaction on such a charge. At the same time one does find a finite Poynting flux at large distances, i.e., for $r\rightarrow \infty$, which is interpreted as power radiated away by the charge 
\cite{5,10}. We, however, have shown from detailed calculations that this Poynting flux actually represents the changing energy in the self-fields of the charge, in synchronism with its changing velocity because of the uniformly acceleration. 
At $t>0$, when the charge is getting accelerated, the Poynting flux at all distances is radially outward from the instantaneous charge position, while for $t<0$, when the charge is getting decelerated, the Poynting flux at all distances is radially inward toward the charge, again at all distances from it. This is consistent with the fact that during the acceleration, the energy in the self-fields is increasing, while during the deceleration, the energy in the self-fields is decreasing. The rate of energy being ``fed'' into fields during the acceleration phase is exactly equal to that of the energy being ``retrieved'' from the fields during the deceleration phase. All this is true even at far-off points from the time retarded positions of the charge. The Poynting flux is nil everywhere at $t=0$, when the charge is momentarily stationary.

It might here be pointed out that the electromagnetic field of a rapidly moving charge with a uniform velocity, $v_0 \rightarrow c$, as seen in Fig. 10 (c), appears to be like the field of a plane wave moving along with the charge, and has for some purposes in the literature been called as radiation comprising virtual photons \cite{1,2}. One might, similarly, consider the electromagnetic field of a uniformly accelerated charge having an instantaneous velocity  $v_0 \rightarrow c$, seen in Fig. 10 (b), to be a radiation comprising of virtual photons moving along with the charge, but there is no electromagnetic radiation `emitted away in a real sense', from a uniformly accelerated charge, in the same way as there is no real radiation emitted away from a charge moving with a uniform velocity.

Actually the field energy is all around the instantaneous position of the charge (falling as $1/R^2$) in regions close to it, while the charge, due to a constant acceleration, may itself be moving to a large distance from the initial position at the retarded time, thus as $r=ct \rightarrow \infty$, the instantaneous charge position, $z_{\rm e}=(z_0^2+c^2t^2)^{1/2} \rightarrow \infty$ too. Poynting flux computed for large $r$, and interpreted in literature as radiation having been emitted by the uniformly accelerating charge, is nothing but (i) that due to the convective flow of the self-fields of the charge in the direction of the present velocity $v_0$ of the charge and (ii) that due to the change taking place in self-field energy due to changing velocity of the charge due to acceleration. Both these quantities will be positive (along +ve $z$-direction and radially outward at large $r=ct$ as the charge will necessarily be in the {\em accelerating} phase at that time while both these quantities will be -ve  (along $-z$-direction as well as a radially inward flow) if one were computing the Poynting flux at $t<0$. 

Since the nature of the uniformly accelerated charge problem is so set up that $r \rightarrow \infty$ always implies $t \rightarrow \infty$, which necessarily means $t>0$ with the charge at that time in the accelerating  phase (and not in a decelerating phase which happens in this set up only at $t<0$), and accordingly velocity $v_0 \rightarrow c$ with $\gamma_0 \rightarrow \infty$, no wonder there is always be an outgoing Poynting flux at $r \rightarrow \infty$. The same will not be true if one were making the calculations by choosing the retarded time as $t \rightarrow -\infty$, so as to have a large $r$ but for $t<0$, then one would always find a radially {\em inward flow}. 

An assertion has, instead, been made that the radiation should be defined by the total rate of energy emitted by the charge at the retarded time $t'$, and is to be calculated by integrating over the surface of the light sphere in the limit of infinite $r=c(t-t')$ for a {\em fixed emission time} $t'$, with both $t\rightarrow\infty$ and $r\rightarrow\infty$ \cite{5,5a}. In fact, the radiation so defined, strictly speaking, may not be in tune with Green's retarded time solution, and could sometime lead to wrong conclusions, especially in the limit $r\rightarrow\infty$ \cite{90}. As has been pointed out, the choice of a {\em fixed emission time} $t'$ to define `radiation' \cite{5,5a}, does nothing but to make the contribution of  the velocity fields to the Poynting flow at a large enough $r$, negligible \cite{90}, while for a uniformly accelerated charge, the contribution of the velocity fields to the Poynting flow cannot be ignored which, matches the contribution of acceleration fields at all distances from the charge, because $v(t')\propto -\dot{v}r/c$. 

At any time, one has to, in fact, take a holistic view of the fields everywhere, that is for all distances from the instantaneous position of the charge, even if the field at various points gets determined from different, time-retarded positions of the charge. One regularly comes across such a situation in case of a charge moving with a uniform velocity, where electric field everywhere is in radial directions from the instantaneous position of the charge, even though the field at different points is determined there too based on different, time-retarded positions of the charge. The argument used there is that in the expression for velocity fields (first term within the square brackets in Eq.~(\ref{eq:38aa}) where no acceleration is involved), the velocity information that gets fed into the field computation is the value at the retarded time and as that value remains constant for the charge moving with a uniform velocity, one is led to the equivalent expression Eq.~(\ref{eq:38ab4}), where the field is radial with respect to the instantaneous position of the charge. The situation is similar in the case of a uniform acceleration case, where the field expression (Eq.~(\ref{eq:38aa})) determines future fields according to the information of the velocity and acceleration of the charge at the retarded time, the field expressions therefore are able to determine the field energy in accordance with the extrapolated velocity at a future time, as long as the acceleration does not change from its value at the retarded time, viz. for a uniform acceleration case. Of course, such a thing is not possible in a time-varying acceleration case, since there is no term in the field expressions for a rate of change of acceleration (or any higher derivatives).

There is thus no radiation going away from the uniformly accelerated charge, instead the fields, including  contribution of the acceleration fields as well, extend from the instantaneous location of the charge, are thereby attached to it, and  not dissociated from the charge as long as it continues to move with the uniform acceleration. 
Naturally there is no radiation reaction on the uniformly accelerated charge, since no field energy is being `radiated away' from such a charge. And, of course, contrary to some earlier opinions \cite{5,PA02}, this also makes it fully conversant with the strong principle of equivalence, where a uniformly accelerated charge is equivalent to a charge supported in a static gravitational field, and such a completely time-static system could not be emitting any radiation. Here we must distinguish two sets of observer. An observer also stationary in the gravitation field is equivalent to the comoving observer in the uniformly accelerated frame. In this case of course everything is static, so no wonder there is no radiation observed. On the other hand, for an observer in free fall in the uniform gravitation field, the charge stationary in the gravitational field by principle of equivalence is equivalent to a charge accelerated uniformly with respect to observer's inertial frame. Such an observer, as we have shown earlier, will observe a finite Poynting flux $\propto v_0^2$ (Eq.~(\ref{eq:38ca})), where  
$v_0$ is the 'present' velocity of the charge with respect to the observer, and thus equal to the Poynting flux that would be seen even for a  charge moving with a uniform velocity $v_0$, which by no means could be called radiation.


The final picture of radiation that emerges is this. In the case of a uniformly accelerated charge, the instantaneous velocity of the charge and the corresponding Lorentz factor $\gamma$, are constantly changing, 
Since the information about the velocity and the acceleration of the charge is contained in the electric field expressions, from the value of acceleration the future velocity of the charge at a time $t$ gets determined, and the acceleration fields in collusion with velocity fields tend to keep the energy in the causally linked spherical surface at $r=ct$ synchronized with the extrapolated velocity of the uniformly accelerated charge at time $t$. As long as the charge continues to move with the same value of (uniform!) acceleration, the things remain synchronized as far as the total energy in the self-fields at time $t$ is concerned. Only when there may be  a change in the acceleration, and since there is no term in the field expressions about the rate of change of acceleration, then the energy in fields does not remain synchronized with that of the actual motion of the charge, which in turn is determined by the actual (changing!) acceleration, resulting in a mismatch between field energy and the actual motion of the charge. As a result any extra energy in fields, no longer part of the self-fields of the charge as determined by its instantaneous motion, is dissociated from the charge, and appears thereby as radiated away energy.

It has sometimes been claimed that the uniformly accelerated charge emits radiation, but a co-accelerating observer cannot see it as the energy emitted from the uniformly accelerated charge goes beyond the horizon, in regions of space-time inaccessible to observers co-accelerating with charge \cite{10,PA02,AL06}.
Actually, the only Poynting flux that goes beyond the horizon is the one in certain $\delta$-fields, arising from and thereby causally related to the charge during its uniform velocity before an acceleration was imposed at an infinite past. It has been explicitly demonstrated \cite{88} that all the energy that goes into $\delta$-fields is neatly explained by the radiation losses, at a rate $\propto -\gamma {\bf v}\cdot\dot{\bf a}$ \cite{68a},
owing to the Abraham-Lorentz radiation reaction \cite{abr05,16,24,3,20,68b}, because of {\em a rate of change of acceleration} the charge, that previously was moving with a uniform velocity, undergoes at that event, thereby neatly explaining the total energy lost by the charge into $\delta$-fields during a transition from a uniform velocity phase to the uniform acceleration phase at infinite past \cite{88}.

A question that still remains to be answered is: why does the observer in the comoving acceleration frame not see a Poynting flux anywhere \cite{PA02}, while the observers in all other inertial frames do find Poynting flux ultimately at large $r$? The answer can be understood this way. Since the observations made in the comoving acceleration frame by an observer are exactly the same as those in the momentarily coincident instantaneous inertial rest frame of the observer, 
and different inertial frames in succession coincide with the comoving acceleration frame at different moments, an observer in the comoving acceleration frame is effectively making observations always from its current instantaneous inertial rest frame, therefore such an observer never sees any Poynting flux which, being $\propto v_0^2$ (Eq.~(\ref{eq:38ca})), is zero in the instantaneous inertial rest frame. In other words, comoving observers do not see any Poynting flow because for them the charge along with its field is static, and the static observers do not detect any Poynting flux from a totally time-static system.

\section{Conclusions}
We demonstrated that for a uniformly accelerated charge, one cannot ignore the velocity fields, which can cancel the acceleration fields even in the far zone. We also showed that while during the acceleration phase, as the velocity of the charge increases, the energy in its self-fields also increases, but during the deceleration phase, as its velocity reduces, so does the energy in its self-fields. This is evident from the direction of the radial component of the Poynting flow, which may though be everywhere outward from the instantaneous position of the charge during its acceleration, implying augmentation of the energy in the fields, yet during the deceleration phase, radial component of the Poynting flow everywhere is along an inward direction toward the charge, indicating a depletion of energy from the fields. Further, it was  shown that in the case of a uniformly accelerated charge, electromagnetic fields, including the acceleration fields, even when at large distances from the time-retarded position of the charge, continue to be all around the instantaneous position of the charge, which itself is moving to infinity due to the constant acceleration. As the uniformly accelerated charge picks up speed so as to approach the speed of light, its fields increasingly resemble those of a charge moving with an equivalent uniform relativistic velocity, where the field strength falls  with square of distance from the current location of the charge, with field lines concentrated along a plane normal to the direction of motion, and no radiation being emitted in either case. Consequently, a uniformly accelerated charge experiences no radiation reaction, a well-known fact. This conclusion is also conversant with the strong principle of equivalence where a uniformly accelerated charge is equivalent to a charge supported in a static gravitational field, and such a completely time-static system could not be emitting any radiation.

\end{document}